\documentclass[sigconf,screen=true, review=false, authordraft=False]{acmart}

\AtBeginDocument{%
  \providecommand\BibTeX{{%
    \normalfont B\kern-0.5em{\scshape i\kern-0.25em b}\kern-0.8em\TeX}}}

\setcopyright{rightsretained}
\copyrightyear{2024}
\acmYear{2024}
\acmDOI{}

\acmConference[EvalAI Workshop at CHI '24]{Make sure to enter the correct
  conference title from your rights confirmation emai}{May 12,
  2024}{Honolulu, HI}
\acmBooktitle{Workshop on Evaluating AI at the CHI conference on Human Factors in Computing Systems,
 May 12,
  2024, Honolulu, HI} 
\acmISBN{}

\begin{document}

\title{Domain-Specific Evaluation Strategies for AI in Journalism}

\author{Sachita Nishal}
\authornote{Both authors contributed equally to this research.}
\email{nishal@u.northwestern.edu}
\affiliation{%
  \institution{Northwestern University}
  \country{USA}
}

\author{Charlotte Li}
\authornotemark[1]
\email{charlotte.li@u.northwestern.edu}
\affiliation{%
  \institution{Northwestern University}
  \country{USA}
}

\author{Nicholas Diakopoulos}
\email{nad@northwestern.edu}
\affiliation{%
 \institution{Northwestern University}
 \country{USA}}






\maketitle

\section{Motivation}

News organizations today rely on AI tools to increase efficiency and productivity across various tasks in news production and distribution. These tools are oriented towards stakeholders such as reporters, editors, and readers. However, practitioners also express reservations around adopting AI technologies into the newsroom, due to the technical and ethical challenges involved in evaluating AI technology and its return on investments \cite{beckettGeneratingChangeGlobal2023}. This is to some extent a result of the lack of domain-specific strategies to evaluate AI models and applications. In this paper, we consider different aspects of AI evaluation (model outputs, interaction, and ethics) that can benefit from domain-specific tailoring, and suggest examples of how journalistic considerations can lead to specialized metrics or strategies. In doing so, we lay out a potential framework to guide AI evaluation in journalism, such as seen in other disciplines \cite{healthdomain, guha2023legalbench}. We also consider directions for future work, as well as how our approach might generalize to other domains.	

\section{Existing AI Evaluation Approaches}

\subsection{From Benchmarking to Human-Centered Evaluation}

Present strategies of evaluating AI models and applications range from generalized quantitative evaluation on benchmark datasets, to scoped qualitative or mixed-methods evaluation in specific contexts. In the former approach, model performance is measured on human-validated metrics, over particular benchmark datasets e.g., for image captioning \cite{chenMicrosoftCOCOCaptions2015}, code synthesis \cite{chenEvaluatingLargeLanguage2021}, action detection \cite{sevillalaraOnlyTimeCan2019}, and so on. While these evaluations can be conducted rapidly and at scale, the metrics and benchmark datasets themselves capture model performance over \textit{generalized notions of "quality" and decomposed tasks}, which limits the their capability for measuring models performance in real-world scenarios \cite{rajiAIEverythingWhole2021}. 

On the other end of this spectrum, human-centered and HCI-based approaches rely on \textit{situated and contextual evaluation} of AI models and applications, such as via highly-scoped user studies (e.g. \cite{xuChatGPTVsGoogle2023}), longitudinal studies (e.g. \cite{longNotJustNovelty2024}), and human-grounded metrics (e.g. \cite{hoffmanMetricsExplainableAI2019}). Recent work illustrates the importance of such scoped approaches, showing how AI models that rank lower on benchmark datasets still perform well in user studies within interactive applications \cite{leeEvaluatingHumanLanguageModel2023}. While these processes allow for more nuanced evaluation in light of a particular application or context, they can be difficult to conduct in a manner that is continuous, iterative, and at scale, which would help to keep pace with model releases, and ensuing novel interactional affordances or ethical issues \cite{etzioniAIAssistedEthics2016}. 

Frameworks for evaluating efficacy of AI models and applications within a specific \textit{domain} (e.g. journalism, medicine, law) can help strike a balance between these approaches. For instance, domain-specific frameworks can guide crafting and validation of domain-specific task benchmarks (e.g. measuring not just "coherence" or "readability" \cite{howcroftTwentyYearsConfusion2020} but "newsworthiness" of LLM-generated news summaries) and draw on domain-specific ethics and values for conducting ethics-based evaluations and audits (e.g. operationalizing and auditing for professional values like "immediacy" in journalism \cite{deuzeWhatJournalismProfessional2005}). To this end, frameworks must identify domain-specific aspects, such as tasks, values, and stakeholder needs, that benchmarking must be scoped toward. They must also provide actionable guidance for continuous evaluation in real-world settings (e.g. newsrooms, hospitals). The next section highlights how this has been approached in prior work, and how journalism could benefit from such domain-specific AI evaluation frameworks.

\subsection{Domain-specific Frameworks for Evaluation}

Prior work on developing domain-specific evaluations of AI mainly exists in the context of healthcare \cite{10.1001/jamanetworkopen.2018.2658,healthdomain,healthcarechecklist} and law \cite{guha2023legalbench}. Similar to journalism, medical and law practitioners' concerns for the deployment of AI into real-world use cases stem from the lack of metrics for evaluating domain-specific quality and ethical alignment. In the medical domain, researchers tackle this issue by proposing a framework that can be incorporated in various stages of model development and deployment, with an emphasis on assessing ethical dimensions of models including privacy, non-maleficence, and explainability \cite{healthdomain}. While parts of this framework can be applicable to journalism, use-cases of AI and ethical concerns in journalism differ from medicine given the public nature and the scope of potential harm caused, calling for the development of journalism-specific frameworks. Such frameworks for benchmarking can supplement qualitative evaluation methods, since they can capture certain aspects of real-world usage scenarios while allowing for iteration and suggesting potential directions for re-design.

The Partnership on AI (PAI) has offered resources to guide AI procurement and use in newsroom, which organize and categorize useful AI tools, and suggest different ways of measuring the outcomes they produce in deployment \cite{sosaPAISeeksPublic2023}. Our work builds in this direction, but offers more specific guidelines on evaluating AI for journalism for both researchers and practitioners, and with additional focus on human-AI interaction as a site of evaluation.

\section{Blueprints for AI Evaluation in Journalism}

This section presents three considerations that evaluations of AI in journalism can include and operationalize: (1) \textit{quality of model outputs}, based on editorial goals and news values (2) \textit{quality of interaction with AI applications}, based on needs and work processes of stakeholders (3) \textit{ethical alignment}, based on professional values and newsroom standards. A useful framework would support practitioners in evaluating AI models and applications along these dimensions, in a manner that is flexible, iterative, and provides feedback for future designs. To actualize these domain-specific metrics, we believe methodologies that invite the collaboration between practitioners and researchers, such as co-design and participatory design, are necessary.

\subsection{Quality of Model Outputs}
\label{blueprint-quality}

To evaluate the quality of AI model outputs, a suite of automatic evaluation metrics and human assessments have been proposed \cite{Maslej2023}. While some of these automatic evaluation metrics can be applied to assess output quality for specific journalistic tasks such as text summarization \cite{lin-2004-rouge, krishna-etal-2023-usb}, machine translation \cite{10.3115/1073083.1073135, goyal-etal-2022-flores}, and object detection \cite{DBLP:journals/corr/LinMBHPRDZ14}, many other journalistic tasks would benefit from evaluation strategies that center more on human assessments. These human assessment evaluation methods are often generalized based on the modality of generation but are not domain-specific. For example, for generated text, human assessments tend to focus on clarity, fluency, accuracy, and coherence of the output \cite{howcroftTwentyYearsConfusion2020}. While these criteria translate well into the domain of journalism, they overlook nuances specific to journalistic writing practice or context, such as specificity or the provision of adequate context. Similarly, for image generation, existing metrics focus mainly on image fidelity, alignment, and counting \cite{10.48550/arxiv.2211.12112, dinh2022tise, saharia2022photorealistic}, with a few exceptions that look at social biases, robustness, and generalization \cite{Cho_2023_ICCV, 10.1109/iccv51070.2023.01834} but lack consideration of editorial judgments around e.g. image framing.

To tailor evaluation strategies towards journalistic uses of AI, researchers and practitioners can draw from news values that guide editorial decision-making. The definitions of news values remain fluid and subjective \cite{parks/newsvalue}, but some of these can be evaluated in model outputs, for instance, in the case of AI-generated news summaries. These elements include \textbf{controversy}, \textbf{surprise}, \textbf{timeliness}, \textbf{negative or positive overtones}, and news organization's \textbf{agenda} \cite{10.1080/1461670x.2016.1150193}. Additionally, the ability of AI tools to support creativity is also important for reporters. Examples of creativity support includes producing \textbf{varied} but relevant outputs, and maintaining professional \textbf{tonality}.

\subsection{Interactions with AI Systems}
\label{blueprint-interaction}

Recent work has called attention to different aspects of \textit{in-situ interactions} with AI systems that may lend themselves well to automated evaluation metrics. These aspects include the \textbf{ease}, \textbf{enjoyment}, and \textbf{feelings of ownership} that users experience when engaged in tasks like question-answering, solving crossword puzzles, and generating metaphors with AI \cite{leeEvaluatingHumanLanguageModel2023}. Human-AI interaction can also be evaluated by the \textit{long-term goals} that a system facilitates, such as promoting \textbf{personal growth} and \textbf{emotional resilience} for users \cite{caiAntagonisticAI2024}; making connections outside users' comfort zone \cite{knijnenburgRecommenderSystemsSelfActualization2016}; and allowing \textbf{customization} of applications or \textbf{appropriation} for novel use-cases \cite{amershiGuidelinesHumanAIInteraction2019, stevensAppropriationInfrastructureMediating2010a}. 

Within journalism, the in-situ interactions and long-term processes that a system could optimize will vary by stakeholders and the tasks facilitated by AI. For instance, AI systems that provide writing feedback to reporters may be evaluated based on the \textbf{new perspectives or angles} they add to a reporter's writing \cite{mottaAnalysisDesignComputational2020} (e.g. by measuring semantic similarity between initial and post-feedback writing). How interactions with AI are configured also impacts the choice of metrics: minor errors in AI outputs (i.e. low \textbf{accuracy}) may be more acceptable to reporters engaged in brainstorming and ideation with text or image models \cite{nishalUnderstandingPracticesComputational2023}, than to readers who only view a final product. Recent work also suggests a desire for long-term \textbf{skill development} among reporters who use AI systems \cite{nishalUnderstandingPracticesComputational2023}, which could be evaluated based on periodic user surveys and AI usage analytics over time. A shared criterion of interaction quality across stakeholders could be the \textbf{enjoyment} they experience when engaged with an AI system, which applies both to reporters who use specific tools in news production (e.g. \cite{thurmanWhenReportersGet2017, reteguiMetricsWorkCase2021}, or to readers who rely on AI (e.g., when solving crossword puzzles \cite{leeEvaluatingHumanLanguageModel2023}). 

Datasets to evaluate these metrics will derive from users' interactions with AI systems, rather than solely model outputs like in Section \ref{blueprint-quality}. Understanding the short and long-term goals of different stakeholders, as well as the configurations of their interactions with AI, can support the design of such interaction metrics. 

\subsection{Ethical Alignment of AI Systems}

Different professions and institutions adhere to different ethical principles and codes of conduct \cite{kovachElementsJournalismWhat2021, deuzeWhatJournalismProfessional2005}. Recent work suggests that subjective and multivalent \textbf{principles of journalistic practice} (e.g. truth, independence, accountability) can be translated to AI systems via value-sensitive design approaches and ethics-based audits \cite{diakopoulosLeveragingProfessionalEthics2024}. Evaluation practices can also measure adherence to \textbf{codes of conduct and style guides} of different newsrooms to be more sensitive to institutional needs (e.g. \cite{thenewyorktimesEthicalJournalism2018, reutersnewsagencyStandardsValuesStyle2021, frokeAssociatedPressStylebook2022}). This translational exercise can also provide practitioners with an opportunity to reflect on the inherent limitations of their existing codes and guides \cite{bien-aimeAPStylebookNormalizes2016, bessetteAssociatedPressStylebook2022}.

Challenges to general-purpose evaluations of ethics and values hold for domain-specific evaluations as well: operationalizing and evaluating ethics can be difficult because generative AI models cannot produce consistent, causal explanations for their outputs \cite{liaoAITransparencyAge2023}. Fine-tuned versions of the same model can also exhibit drastic variations in their adherence to ethical values. Journalism's orientation toward timeliness of communication also exacerbates these challenges. This further indicates the need for iterative and continuous evaluations of AI models and applications \cite{mokanderEthicsBasedAuditingAutomated2021}. It also indicates a need for technical innovation to support operationalization and of journalism-specific ethics and values in AI systems.

\section{Future Directions and Conclusion}

In this position paper, we summarized current approaches for AI evaluation, and made recommendations for additional journalism-specific evaluation criteria across three different aspects of AI evaluation. The primary goal of this paper is to call for researchers and practitioners to come together and develop domain-specific frameworks for evaluating AI systems, so the adaptation of AI tools into newsrooms become easier and more equitable. Such frameworks can enable the development of procurement guidelines for AI tools in newsrooms, as seen in prior work \cite{sosaPAISeeksPublic2023}. We further hope this approach can empower stakeholders to create newsroom-specific custom evaluation datasets, for both short-term and longitudinal assessments of the technology.

\bibliographystyle{ACM-Reference-Format}
\bibliography{project-bib}



\end{document}